\documentclass[preprint,showpacs,preprintnumbers,amsmath,amssymb]{revtex4}

\usepackage{graphicx}
\usepackage{dcolumn}
\usepackage{bm}

\usepackage[english]{babel}
\usepackage{amsmath,amsfonts,amssymb}
  \usepackage{epsfig}
\usepackage{axodraw}

\allowdisplaybreaks

\begin{document}


\title{Electromagnetic Corrections \\
to the \(\eta\rightarrow 3\pi\) Neutral Decay}

\author{A.  Nehme}
\email{anehme@ul.edu.lb}
\affiliation{%
Lebanese University, \\
Faculty of Sciences at Nabatieh, Lebanon}%

\author{S. Zein}
\affiliation{%
Universit\'e Libanaise, \\
Ecole Doctorale de Sciences et Technologies, \\
B. P. 5, Hadath, Liban}

\date{\today}

\begin{abstract}
Sutherland's theorem dictates that the contribution of the electromagnetic interaction to the decay process \(\eta\rightarrow 3\pi^{0}\) is neglected with respect to the one coming from the difference between the up and down quark
masses. In the framework of chiral perturbation theory including virtual
photons, we calculated the main diagram concerning the exchange of a virtual
photon between two intermediate charged pions. The correction induced by
this diagram on the slope parameter amounts to \(17\%\) of the correction
induced by the pure strong interaction at one-loop level. If this result
is maintained when considering all the diagrams at the chiral order we are
working, we can say without any doubt that Sutherland's theorem is strongly
violated. As a direct consequence, any determination of light quark masses
from the present decay \textit{should} take into account the electromagnetic
interaction.
\end{abstract}

\maketitle

\section{\label{sec:intro}Introduction}

The decay \(\eta\rightarrow\pi^{0}\pi^{0}\pi^{0}\) is forbidden by isospin
symmetry of the strong interaction. This symmetry is explicitly broken by
electromagnetism on the one hand, and by the difference between up and down
quark masses on the other hand. The decay amplitude takes then contributions from the electromagnetic fine structure constant \(\alpha_{\text{fs}}\) and \(m_{d}-m_{u}\,.\) Sutherland's theorem states that the electromagnetic contribution is suppressed with respect to the other one \cite{S66,BS68}. As a consequence, the decay \(\eta\rightarrow 3\pi\)
has been considered over the last thirty years as the golden process for the determination of the difference \(m_{d}-m_{u}\)
by comparing the theoretical predictions with the experimental measurements
(See \cite{KKNZ11} and references therein).
However, the data for the slope in the corresponding Dalitz plot
show a big discrepancy between the experimental value and the theoretical
predictions based on chiral perturbation theory \cite{BG07}. This fact motivated
the scientific community to perform very accurate measurements at the experimental
level \cite{Prakhov:2008ff,Adolph:2008vn,Ambrosinod:2010mj}. On the other
side, several attempts have been conducted at the theoretical level. For
instance, a model-independent study resting only on chiral symmetry and experimental
data was performed in \cite{DNT08}. The authors of \cite{KKNZ11} used another approach based on data and analytic dispersive representation including first-order
isospin breaking effects. Rescattering effects were considered in \cite{Schneider:2010hs}
through a modified non relativistic effective field theory beyond one loop
including isospin-breaking corrections. It was claimed in the preceding citation
that ``the effect of photon exchange inside the charged pion loops on the Dalitz plot expansion is small, even on the scale of the other small isospin-breaking
effects''.  We believe, and will partially prove, that this is not the case.
Our statement rests on the observation that the re-scattering effects between two pions in the final state is dominated by the Coulomb interaction which,
in two different (but related) processes, gives a relatively valuable correction\cite{Knecht:1997jw,Knecht:2002gz,Nehme:2004xy}. Motivated by this observation, we calculate in the present work the correction
induced on the slope parameter of the decay \(\eta\rightarrow 3\pi^{0}\)
by photon exchange between two intermediate charged pions. To do so, the
corresponding Feynman integral has been reduced to a set of five Master Integrals
following Tarasov's reduction algorithm \cite{T96,T97}, which is implemented
in computer algebraic systems as a Mathematica package called Tarcer \cite{Mertig:1998vk}.
The Master Integrals are then obtained in dimensional regularization by solving hypergeometric differential equations as reviewed in \cite{Argeri:2007up}.
As a next step, we expand the decay amplitude around its value at the center
of the Dalitz plot and derive an analytic expression for the slope parameter.
The value of the slope is then deduced in four dimensions by expanding around
\(D=4\). We finally comment on our finding.

\section{\label{amp}The decay amplitude}

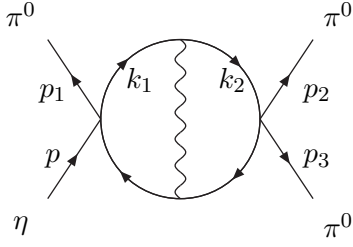
\begin{figure}[ht]
\begin{center}
\begin{picture}(120,80)(0,0)
\ArrowLine(10,10)(30,40)
\ArrowLine(30,40)(10,70)
\ArrowLine(90,40)(110,10)
\ArrowLine(90,40)(110,70)
\ArrowArcn(60,40)(30,180,90)
\ArrowArcn(60,40)(30,90,0)
\ArrowArcn(60,40)(30,0,90)
\ArrowArcn(60,40)(30,90,180)
\Photon(60,70)(60,10){2}{6}
\Text(0,0)[]{$\eta$}
\Text(0,80)[]{$\pi^{0}$}
\Text(120,0)[]{$\pi^{0}$}
\Text(120,80)[]{$\pi^{0}$}
\Text(12,25)[]{$p$}
\Text(12,50)[]{$p_{1}$}
\Text(112,50)[]{$p_{2}$}
\Text(112,25)[]{$p_{3}$}
\Text(80,55)[]{$k_{2}$}
\Text(45,55)[]{$k_{1}$}
\end{picture}
\caption{\label{fig:main_diagram}In this diagram \(q\equiv p-p_{1}=p_{2}+p_{3}\).}
\end{center}
\end{figure}

The dynamics of the process
\begin{equation}
\eta (p)\longrightarrow\pi^{0}(p_{1})+\pi^{0}(p_{2})+\pi^{0}(p_{3})
\end{equation}
is studied in terms of the Mandelstam variables
\begin{equation}
s=(p-p_{1})^{2}\,, \quad t=(p-p_{2})^{2}\,, \quad u=(p-p_{3})^{2}\,,
\end{equation}
subject to the constraint
\begin{equation}
s+t+u=M_{\eta}^{2}+3M_{\pi^{0}}^{2}\equiv 3s_{0}\,.
\end{equation}
The decay amplitude \(\mathcal{M}\) is defined as
\begin{equation}
\left\langle \pi^{0}\pi^{0}\pi^{0}\right\vert\eta\rangle =i(2\pi )^{4}\delta^{4}(p_{1}+p_{2}+p_{3}-p)\mathcal{M}\,,
\end{equation}
and can be written by symmetry considerations like
\begin{equation}
\mathcal{M}(s,t,u)=M(s)+M(t)+M(u)\,.
\end{equation}
In the framework of chiral perturbation theory, we write the \(s\)-channel amplitude as
\begin{equation}
M(s)=-\frac{B_{0}(m_{d}-m_{u})}{3\sqrt{3}F_{\pi}^{2}}\left[ 1+\delta_{\text{str}}(s)+\delta_{\text{em}}(s)+\delta_{\gamma}(s) \right] +\tilde \delta_{\text{em}}(s)\,,
\end{equation}
where \(\delta_{\text{str}}\) is of \(\mathcal{O}(p^{2})\), \(\delta_{\text{em}}\)
of \(\mathcal{O}(e^{2})\), and \(\tilde \delta_{\text{em}}\) of \(\mathcal{O}(e^{2}p^{2})\)
in the chiral counting and have been calculated in \cite{GL85}, \cite{BKW96},
and \cite{DNT08}, respectively. The correction \(\delta_{\gamma}\) is of
\(\mathcal{O}(e^{2}p^{2})\) and is supposed to be small compared to the others.
The main diagram contributing to \(\delta_{\gamma}\) is sketched in Fig.~\ref{fig:main_diagram}.
Working in the framework of chiral perturbation theory including photons
\cite{Urech:1994hd},
we find that, up to polynomials in the kinematic invariants, the main diagram is given in terms of \(D\)-dimensional integrals of the form
\begin{equation}\label{eq:main_integral}
\frac{1}{\pi^{D}}\iint\frac{d^{D}k_{1}d^{D}k_{2}(k_{1}^{2})^{i}(k_{2}^{2})^{j}(q\cdot
k_{1})^{k}(q\cdot k_{2})^{l}(k_{1}\cdot k_{2})^{m}}{(k_{1}^{2}-M_{\pi}^{2})(k_{2}^{2}-M_{\pi}^{2})[(k_{1}-q)^{2}-M_{\pi}^{2}][(k_{2}-q)^{2}-M_{\pi}^{2}](k_{1}-k_{2})^{2}}\,.
\end{equation}
Applying Tarasov's reduction algorithm,
integrals (\ref{eq:main_integral}) are reduced to a set of five Master Integrals,
\(J_{1}\), \(J_{2}\), \(JT\), \(J^{2}\), and \(T^{2}\) with
\begin{eqnarray}
T
&=& \frac{1}{\pi^{D/2}}\int\frac{d^{D}k}{k^{2}-M_{\pi}^{2}}\,, \\
J
&=& \frac{1}{\pi^{D/2}}\int\frac{d^{D}k}{(k^{2}-M_{\pi}^{2})[(k-q)^{2}-M_{\pi}^{2}]}\,,
\\
J_{1}
&=& \frac{1}{\pi^{D}}\iint\frac{d^{D}k_{1}d^{D}k_{2}}{(k_{2}^{2}-M_{\pi}^{2})[(k_{1}-q)^{2}-M_{\pi}^{2}](k_{1}-k_{2})^{2}}\,,
\\
J_{2}
&=& \frac{1}{\pi^{D}}\iint\frac{d^{D}k_{1}d^{D}k_{2}(k_{1}\cdot k_{2})}{(k_{2}^{2}-M_{\pi}^{2})[(k_{1}-q)^{2}-M_{\pi}^{2}](k_{1}-k_{2})^{2}}\,.
\end{eqnarray}
We found that the contribution of the main diagram to \(\delta_{\gamma}\)
takes the following form in \(D\) dimensions
\begin{eqnarray}
\delta_{\gamma}
&=& -\frac{\alpha_{\text{fs}}}{3F_{\pi}^{2}}\frac{1}{M_{\pi^{0}}^{2}(M_{\eta}^{2}-M_{\pi^{0}}^{2})}\frac{1}{s^{3}\sigma^{4}(s)}\times
\nonumber\\
&& \frac{1}{(D-4)(D-3)}\frac{1}{(4\pi
)^{D-1}}(c_{1}J_{1}+c_{2}J_{2}+c_{3}JT+c_{4}J^{2}+c_{5}T^{2})\,,
\end{eqnarray}
where
\begin{equation}
\sigma (s)=\sqrt{1-\frac{4M_{\pi}^{2}}{s}}\,,
\end{equation}
and,
\begin{eqnarray}
c_{1}
&=& -(-384 D^2 M_{\pi }^{10}+1536 D M_{\pi }^{10}-1152 M_{\pi }^{10}+608 D^2 s M_{\pi }^8-2528 D s M_{\pi }^8\nonumber\\
&& +864 s M_{\pi }^8+184 D^2 s^2 M_{\pi }^6-624 D s^2 M_{\pi }^6+2240 s^2 M_{\pi }^6-650 D^2 s^3 M_{\pi }^4+2810 D s^3 M_{\pi }^4\nonumber\\
&& -3408 s^3 M_{\pi }^4+257 D^2 s^4 M_{\pi }^2-1299 D s^4 M_{\pi }^2+1564 s^4 M_{\pi }^2+12 D^2 s^5-42 D s^5+48 s^5)\,, \nonumber\\
c_{2}
&=& 2 (-576 D^2 M_{\pi }^8+2880 D M_{\pi }^8-3456 M_{\pi }^8+1360 D^2 s M_{\pi }^6-6880 D s M_{\pi }^6\nonumber\\
&& +7776 s M_{\pi }^6-980 D^2 s^2 M_{\pi }^4+5168 D s^2 M_{\pi }^4-5568 s^2 M_{\pi }^4+187 D^2 s^3 M_{\pi }^2\nonumber\\
&& -1213 D s^3 M_{\pi }^2+1356 s^3 M_{\pi }^2+36 D^2 s^4-126 D s^4+144 s^4)\,, \nonumber\\
c_{3}
&=& -s^{2}\sigma^{2}(s)(56 D^2 M_{\pi }^6-376 D M_{\pi }^6+416 M_{\pi }^6-66 D^2 s M_{\pi }^4+522 D s M_{\pi }^4-600 s M_{\pi }^4\nonumber\\
&& +D^2 s^2 M_{\pi }^2-119 D s^2 M_{\pi }^2+148 s^2 M_{\pi }^2+12 D^2 s^3-42 D s^3+48 s^3)\,, \nonumber\\
c_{4}
&=& 3s^{2}\sigma^{2}(s)M_{\pi}^{2}(3 s-4 M_{\pi }^2)(s-M_{\pi }^2)(D-3)(-4 D M_{\pi }^2+8 M_{\pi }^2+3 D s-8 s)\,, \nonumber\\
c_{5}
&=& -192 D^2 M_{\pi }^8+1152 D M_{\pi }^8-1536 M_{\pi }^8+560 D^2 s M_{\pi }^6-3312 D s M_{\pi }^6\nonumber\\
&& +3712 s M_{\pi }^6-396 D^2 s^2 M_{\pi }^4+2568 D s^2 M_{\pi }^4-2640 s^2 M_{\pi }^4+31 D^2 s^3 M_{\pi }^2\nonumber\\
&& -489 D s^3 M_{\pi }^2+524 s^3 M_{\pi }^2+36 D^2 s^4-126 D s^4+144 s^4\,.
\end{eqnarray}

\section{The Master Integrals}

Three of the Master Integrals are given in terms of one-loop integrals \(T\)
and \(J\),
\begin{eqnarray}
T
&=& -iM_{\pi}^{D-2}\Gamma (1-D/2)\,, \\
J
&=& i\left\{ 1-M_{\pi}^{D-4}\Gamma (1-D/2)+\sigma \left[ \ln\left( \frac{1-\sigma}{1+\sigma} \right) +i\pi \right]\right\}\,.
\end{eqnarray}
The remaining two integrals are found to satisfy the system of
first order differential equations,
\begin{eqnarray}
\frac{dJ_{1}}{ds}
&=& \frac{(5-2D)s+2(3-2D)M_{\pi}^{2}}{s(4M_{\pi}^{2}-s)}J_{1}+\frac{6(D-2)}{s(4M_{\pi}^{2}-s)}J_{2}+\frac{2(D-2)}{s(4M_{\pi}^{2}-s)}T^{2}\,,
\\
\frac{dJ_{2}}{ds}
&=& -\frac{8(D-1)M_{\pi}^{4}+2(3D-8)sM_{\pi}^{2}+(D-2)s^{2}}{4s(4M_{\pi}^{2}-s)}J_{1}
\nonumber\\
&& +\frac{D-2}{2s}\left( \frac{8M_{\pi}^{2}+s}{4M_{\pi}^{2}-s} \right)J_{2}+\frac{D-2}{4s}\left( \frac{4M_{\pi}^{2}+s}{4M_{\pi}^{2}-s} \right)T^{2}\,.
\end{eqnarray}
Such a system is equivalent to a second order differential equation
\begin{eqnarray}
\frac{d^{2}J_{1}}{ds^{2}}
&=& -\frac{(D-2)^{2}}{2s^{2}(4M_{\pi}^{2}-s)}T^{2}-\frac{3(4M_{\pi}^{2}+Ds-4s)}{2s(4M_{\pi}^{2}-s)}\frac{dJ_{1}}{ds} \nonumber\\
&& +\frac{(D-3)[2(D-2)M_{\pi}^{2}+(D-4)s]}{2s^{2}(4M_{\pi}^{2}-s)}J_{1}\,.
\end{eqnarray}
Differentiating once more, we obtain a third order differential equation
\begin{eqnarray}
\frac{d^{3}J_{1}}{ds^{3}}
&=& \frac{(D-4)(D-3)}{2s^{2}(4M_{\pi}^{2}-s)}J_{1}-\frac{28M_{\pi}^{2}+3Ds-18s}{2s(4M_{\pi}^{2}-s)}\frac{d^{2}J_{1}}{ds^{2}} \nonumber\\
&& +\frac{2D^{2}M_{\pi}^{2}+sD^{2}-10DM_{\pi}^{2}-13sD+36s}{2s^{2}(4M_{\pi}^{2}-s)}\frac{dJ_{1}}{ds}\,.
\end{eqnarray}
Making the change of variable,
\begin{equation}
x=-\frac{s}{4M_{\pi}^{2}}\,,
\end{equation}
we find that \(J_{1}\) satisfies the hypergeometric differential equation
\begin{eqnarray}
&& x^{2}(1+x)\frac{d^{3}J_{1}}{dx^{3}}-x\left[ \frac{3(D-6)}{2}x-\frac{7}{2} \right]\frac{d^{2}J_{1}}{dx^{2}} \nonumber\\
&& +\left[ \frac{(D-4)(D-9)}{2}x-\frac{D(D-5)}{4} \right]\frac{dJ_{1}}{dx}+\frac{(3-D)(4-D)}{2}J_{1}=0\,.
\end{eqnarray}
The general solution of such an equation is
\begin{eqnarray}
J_{1}
&=& \phantom{}_{3}F_{2}\left[
\begin{array}{c}
(3-D),(4-D)/2,1\\
D/2,(5-D)/2
\end{array}; -x\right] A \nonumber\\
&& +(-x)^{-(D-2)/2}\phantom{}_{3}F_{2}\left[\begin{array}{c}(8-3D)/2,(3-D),(4-D)/2\\
(4-D)/2,(7-2D)/2
\end{array}; -x\right] B \nonumber\\
&& +(-x)^{(D-3)/2}\phantom{}_{3}F_{2}\left[\begin{array}{c}(3-D),1/2,(D-1)/2\\
(2D-3)/2,(D-1)/2
\end{array}; -x\right] C\,.
\end{eqnarray}
In order to determine the integration constants \(A\), \(B\), and \(C\),
we notice that
\begin{equation}
J_{1}(0)=\frac{1}{2M_{\pi}^{2}}\left( \frac{D-2}{D-3} \right) T^{2}\,,
\end{equation}
is regular for \(2<D<4\). It follows that \(B=C=0\) since the terms \((-x)^{-(D-2)/2}\)
and \((-x)^{(D-3)/2}\) are divergent for \(x=0\) and in the range in question. Moreover, \(_{3}F_{2}\) is equal to one for \(x=0\). This means that \(A=J_{1}(0)\)
and
\begin{equation}
J_{1}=\frac{1}{2M_{\pi}^{2}}\left( \frac{D-2}{D-3} \right)\phantom{}_{3}F_{2}\left[\begin{array}{c}(3-D),(4-D)/2,1\\ D/2,(5-D)/2
\end{array}; -x\right] T^{2}\,.
\end{equation}
Having \(J_{1}\), we obtain \(J_{2}\) from
\begin{equation}
J_{2}=-\frac{1}{3}T^{2}+\frac{2M_{\pi}^{2}}{3(D-2)}x(1+x)\frac{dJ_{1}}{dx}-\frac{M_{\pi}^{2}[2(2D-5)x+3-2D]}{3(D-2)}J_{1}\,,
\end{equation}
by the use of
\begin{eqnarray}
&& \frac{d}{dx}\,_{3}F_{2}\left[\begin{array}{c}(3-D),(4-D)/2,1\\
D/2,(5-D)/2
\end{array}; -x\right]
= \nonumber\\
&& \frac{2(D-3)(D-4)}{D(D-5)}\,_{3}F_{2}\left[\begin{array}{c}(4-D),(6-D)/2,2\\ (D+2)/2,(7-D)/2
\end{array}; -x\right]\,.
\end{eqnarray}
We get
\begin{eqnarray}
J_{2}
&=& \left\{
- \frac{2(2D-5)x+3-2D}{6(D-3)}
\,_{3}F_{2}\left[\begin{array}{c}(3-D),(4-D)/2,1\\
D/2,(5-D)/2
\end{array}; -x\right]\right.\nonumber\\
&& \left.+\frac{2 (D-4)}{3 D (D-5)}x(x+1)
     _{3}F_{2}\left[\begin{array}{c}(4-D),(6-D)/2,2\\
(D+2)/2,(7-D)/2
\end{array}; -x\right]-\frac{1}{3}\right\}T^{2}\,.
\end{eqnarray}

\section{The slope parameter}

The amplitude is expanded around its value at the center of the Dalitz plot
as
\begin{equation}
\left\vert \mathcal{M}(s,t,u) \right\vert^{2}=\left\vert \mathcal{M}(s_{0},s_{0},s_{0}) \right\vert^{2}\left\{ 1+2\alpha z \right\}\,,
\end{equation}
where
\begin{equation}
z=\frac{3}{2M_{\eta}^{2}(M_{\eta}-3M_{\pi^{0}})^{2}}\left\{ (s-s_{0})^{2}+(t-s_{0})^{2}+(u-s_{0})^{2} \right\}\,.
\end{equation}
We write the slope parameter as
\begin{equation}
\alpha =\alpha_{\text{str}}+\alpha_{\text{em}}+\alpha_{\gamma}+\frac{M_{\pi^{\pm}}^{2}-M_{\pi^{0}}^{2}}{B_{0}(m_{d}-m_{u})}\tilde \alpha_{\text{em}}\,.
\end{equation}
The contribution of the main diagram to the slope parameter is
\begin{equation}
\alpha_{\gamma}=\frac{M_{\eta}^{2}}{9}(M_{\eta}-3M_{\pi^{0}})^{2}\,\text{Re}\,\delta_{\gamma}''(s_{0})\,.
\end{equation}
We can express \(\alpha_{\gamma}\) in terms of the five Master Integrals
thanks to the differential equations satisfied by the latter. Up to now,
we have considered only the differential equations satisfied by \(J_{1}\) and \(J_{2}\). It is easy to check that the one-loop integral satisfies the
following differential equation
\begin{equation}
\frac{dJ}{ds}=\frac{(D-4)s+4M_{\pi}^{2}}{2s(s-4M_{\pi}^{2})}J-\frac{D-2}{s(s-4M_{\pi}^{2})}T\,.
\end{equation}
Differentiating \(\delta_{\gamma}\) twice with respect to \(s\) and using
the preceding relation at each step, we obtain
\begin{eqnarray}
\alpha_{\gamma}
&=& -\frac{\alpha_{\text{fs}}}{108F_{\pi}^{2}}\frac{M_{\eta}^{2}(M_{\eta}-3M_{\pi^{0}})^{2}}{M_{\pi^{0}}^{2}(M_{\eta}^{2}-M_{\pi^{0}}^{2})}\frac{1}{s_{0}^{7}\sigma^{8}(s_{0})}\times\nonumber\\
&& \frac{1}{(D-4)(D-3)}\frac{1}{(4\pi
)^{D-1}}\,\text{Re}\left( d_{1}J_{1}+d_{2}J_{2}+d_{3}JT+d_{4}J^{2}+d_{5}T^{2} \right)_{s=s_{0}}\,,
\end{eqnarray}
with
\begin{eqnarray}
d_{1}
&=& 6144(D^4-16 D^3+86 D^2-176 D+105) M_{\pi }^{14}-512(25 D^4-360 D^3+1511 D^2\nonumber\\
&& -2214 D+621) s M_{\pi }^{12}+128 (48 D^4-374 D^3+257 D^2+3240 D-5856) s^2 M_{\pi }^{10}\nonumber\\
&& -32 (50 D^4+2009 D^3-17080 D^2+43918 D-36964) s^3 M_{\pi }^8+8 (397 D^4-684 D^3\nonumber\\
&& -15994 D^2+65826 D-69384) s^4 M_{\pi }^6+2 (201 D^4+6971 D^3-60836 D^2\nonumber\\
&& +161172 D-139248) s^5 M_{\pi }^4+(-1697 D^4+16825 D^3-62006 D^2\nonumber\\
&& +101120 D-61664) s^6 M_{\pi }^2+6 D (-2 D^3+11 D^2-22 D+16) s^7\,, \nonumber\\
d_{2}
&=& 2[-9216 (D^4-17 D^3+101 D^2-247 D+210) M_{\pi }^{12}+256 (103 D^4-1527 D^3\nonumber\\
&& +7880 D^2-17199 D+13446) s M_{\pi }^{10}-64 (454 D^4-5634 D^3+25631 D^2\nonumber\\
&& -50730 D+36744) s^2 M_{\pi }^8+48 (398 D^4-3551 D^3+12214 D^2-19708 D\nonumber\\
&& +12600) s^3 M_{\pi }^6+4 (-2491 D^4+18774 D^3-43814 D^2+26916 D+11088) s^4 M_{\pi }^4\nonumber\\
&& +(2971 D^4-30543 D^3+114482 D^2-188208 D+115296) s^5 M_{\pi }^2\nonumber\\
&& +18 D (2 D^3-11 D^2+22 D-16) s^6]\,, \nonumber\\
d_{3}
&=& s^{2}\sigma^{2}(s)[384 (10 D^3-77 D^2+207 D-172) M_{\pi }^{10}+32 (18 D^4-320 D^3\nonumber\\
&& +1733 D^2-3751 D+2764) s M_{\pi }^8-8 (187 D^4-1741 D^3+5696 D^2\nonumber\\
&& -7898 D+3816) s^2 M_{\pi }^6+6 (209 D^4-1951 D^3+5866 D^2-6400 D\nonumber\\
&& +1568) s^3 M_{\pi }^4+(-325 D^4+3953 D^3-16398 D^2+29008 D\nonumber\\
&& -18848) s^4 M_{\pi }^2+6 D (-2 D^3+11 D^2-22 D+16) s^5]\,, \nonumber\\
d_{4}
&=& -12M_{\pi}^{2}s^{2}\sigma^{2}(s)[512 (D^2-5 D+6) M_{\pi }^{10}+128 (D^3-11 D^2\nonumber\\
&& +36 D-36) s M_{\pi }^8+16 (D^4-16 D^3+91 D^2-216 D+180) s^2 M_{\pi }^6\nonumber\\
&& -8 (5 D^4-38 D^3+92 D^2-65 D-12) s^3 M_{\pi }^4+(D-3)^2 (33 D^2\nonumber\\
&& -68 D-64) s^4 M_{\pi }^2-3 (D-3)^2 (3 D^2-14 D+16) s^5]\,, \nonumber\\
d_{5}
&=& -3072 (D^4-18 D^3+115 D^2-306 D+280) M_{\pi }^{12}+256 (37 D^4-615 D^3+3401 D^2\nonumber\\
&& -7860 D+6444) s M_{\pi }^{10}-64 (183 D^4-2517 D^3+12185 D^2-25152 D+18772) s^2 M_{\pi }^8\nonumber\\
&& +16 (568 D^4-5294 D^3+18567 D^2-30394 D+19784) s^3 M_{\pi }^6+4 (-1386 D^4+10909 D^3\nonumber\\
&& -25796 D^2+15140 D+7808) s^4 M_{\pi }^4+5 (359 D^4-3883 D^3+14986 D^2-25136 D\nonumber\\
&& +15648) s^5 M_{\pi }^2+18 D (2 D^3-11 D^2+22 D-16) s^6\,.
\end{eqnarray}

\section{Results and Conclusions}

We first expand \(\delta_{\gamma}\) around \(D=4\) and then use the following
numerical values
\begin{equation}
\alpha_{\text{fs}} =\frac{1}{137.04}\,, \quad (F_{\pi},M_{\pi^{0}}\equiv M_{\pi},M_{\eta})=(92.42,\,
139.57,\, 547.30)\;\text{MeV}\,.
\end{equation}
We found that the amplitude, \(\delta_{\gamma}\), possesses a pole at the
lower bound of the allowed kinematical region, that is, for \(s=4M_{\pi^{0}}^{2}\).
At the upper bound, \(s=(M_{\eta}-M_{\pi^{0}})^{2}\), the calculated electromagnetic
correction amounts to \(3\)-\(4\%\) of the Tree level amplitude. This points
out that the present correction can not be simply neglected as dictated by
Sutherland's theorem and as is widely believed by physicists.

We next do the same for the slope parameter and obtain
\begin{equation}
\alpha_{\gamma}=0.0029\,.
\end{equation}
This value is to be compared with the one-loop strong and electromagnetic corrections,
\begin{equation}
\alpha_{\text{str}}=0.0179\,, \qquad \alpha_{\text{em}}=-0.0011\,,
\end{equation}
respectively. The first conclusion we draw comes from the comparison between
the electromagnetic corrections at one- and two-loop levels. Although the
two-loop calculation does not include the contribution of all the diagrams,
it is three times bigger than the one-loop calculation. Since the one-loop
correction is originating from the mass difference between charged and neutral
pions, and the two-loop correction considered here is due to photon exchange
between charged intermediate pions, we can claim that the effect of soft
virtual photons is more important than the one for hard virtual photons.
Obviously, this conclusion does no more hold if some cancellations occur
between the various contributions if we consider all diagrams with a photon exchange. This point should constitute a sufficient motivation for calculating
the full amplitude including the exchange of one photon between charged
intermediate light mesons.

We draw the second conclusion by comparing the size of the electromagnetic
correction to the size of the one-loop pure strong correction. If we consider
only the contribution of the photon exchange diagram, we find that the former
amounts to \(17\%\) of the latter. Adding both one- and two-loop corrections,
we find that the electromagnetic interaction contributes about \(11\%\) of
the pure strong interaction at one loop. If we add both strong and electromagnetic
corrections, at the one-loop order for the first, and the two-loop order
for the second, we obtain for the slope parameter
\begin{equation}
\begin{array}{ccccccccc}
\alpha & = & \underbrace{0.0179} & + & \underbrace{-0.0011} & + & \underbrace{0.0029} & = & +0.0198\,, \\
 &  & \mathcal{O}(p^{4}) &  & \mathcal{O}(e^{2}p^{2}) &  & \mathcal{O}(e^{2}p^{4}) &  &  \\
\end{array}
\end{equation}
to be compared with the most recent experimental value \cite{Ambrosinod:2010mj},
\begin{equation}
\alpha_{\text{exp}}=-0.0301\pm 0.0035_{-0.0035}^{+0.0022}\,,
\end{equation}
noting this endless sign discrepancy between the chiral perturbation theory
prediction and observation.

\begin{acknowledgments}
S. Z. wishes to acknowledge the ``Laboratoire de Physique des Mat\'eriaux''
staff for their support and kindness.
\end{acknowledgments}


\end{document}